\documentclass[sigconf]{acmart}
\AtBeginDocument{%
  }

\usepackage{multirow}
\usepackage{enumitem}
\usepackage{colortbl}
\usepackage{tablefootnote}
\usepackage{makecell}
\usepackage{balance}
\usepackage{fontawesome}
\usepackage{algorithm}
\usepackage{algorithmic}

\usepackage{fancyvrb}
\usepackage{xcolor}

\newcommand{\ie}{\emph{i.e., }}
\newcommand{\eg}{\emph{e.g., }}
\newcommand{\etal}{\emph{et al. }}

\newcommand{\wrt}{\emph{w.r.t. }}

\setcopyright{acmlicensed}
\copyrightyear{2018}
\acmYear{2018}
\acmDOI{XXXXXXX.XXXXXXX}
\acmConference[Conference acronym 'XX]{Make sure to enter the correct
  conference title from your rights confirmation email}{June 03--05,
  2018}{Woodstock, NY}
\acmISBN{978-1-4503-XXXX-X/2018/06}

\begin{document}

\title{SemCORE: A Semantic-Enhanced Generative Cross-Modal Retrieval Framework with MLLMs}
\author{Haoxuan Li}
\affiliation{%
  \institution{University of Electronic Science and Technology of China}
  \city{Chengdu}
  \country{China}
}
\email{lhx980610@gmail.com}

\author{Yi Bin}
\affiliation{%
  \institution{Tongji University}
  \city{Shanghai}
  \country{China}
}
\email{yi.bin@hotmail.com}

\author{Yunshan Ma}
\affiliation{%
  \institution{Singapore Management University}
  \country{Singapore}
}
\email{ysma@smu.edu.sg}

\author{Guoqing Wang}
\affiliation{%
  \institution{University of Electronic Science and Technology of China}
  \city{Chengdu}
  \country{China}
}
\email{gqwang0420@uestc.edu.cn}

\author{Yang Yang}
\affiliation{%
  \institution{University of Electronic Science and Technology of China}
  \city{Chengdu}
  \country{China}
}
\email{yang.yang@uestc.edu.cn}

\author{See-Kiong Ng}
\affiliation{%
  \institution{National University of Singapore}
  \country{Singapore}
}
\email{seekiong@nus.edu.sg}

\author{Tat-Seng Chua}
\affiliation{%
  \institution{National University of Singapore}
  \country{Singapore}
}
\email{dcscts@nus.edu.sg}

\renewcommand{\shortauthors}{Haoxuan Li et al.}

\begin{abstract}
  Cross-modal retrieval (CMR) is a fundamental task in multimedia research, focused on retrieving semantically relevant targets across different modalities. While traditional CMR methods match text and image via embedding-based similarity calculations, recent advancements in pre-trained generative models have established generative retrieval as a promising alternative. 
  This paradigm assigns each target a unique identifier and leverages a generative model to directly predict identifiers corresponding to input queries without explicit indexing. Despite its great potential, current generative CMR approaches still face semantic information insufficiency in both identifier construction and generation processes.
  To address these limitations, we propose a novel unified \textbf{Sem}antic-enhanced generative \textbf{C}ross-m\textbf{O}dal \textbf{RE}trieval framework (\textbf{SemCORE}), designed to unleash the semantic understanding capabilities in generative cross-modal retrieval task.
  Specifically, we first construct a Structured natural language IDentifier (SID) that effectively aligns target identifiers with generative models optimized for natural language comprehension and generation. Furthermore, we introduce a Generative Semantic Verification (GSV) strategy enabling fine-grained target discrimination. Additionally, to the best of our knowledge, SemCORE is the first framework to simultaneously consider both text-to-image and image-to-text retrieval tasks within generative cross-modal retrieval. 
  Extensive experiments demonstrate that our framework outperforms state-of-the-art generative cross-modal retrieval methods. Notably, SemCORE achieves substantial improvements across benchmark datasets, with an \textbf{average increase of 8.65 points in Recall@1} for text-to-image retrieval. 
\end{abstract}




\maketitle

\begin{figure}[!t]
\centering
\includegraphics[width=0.92\linewidth]{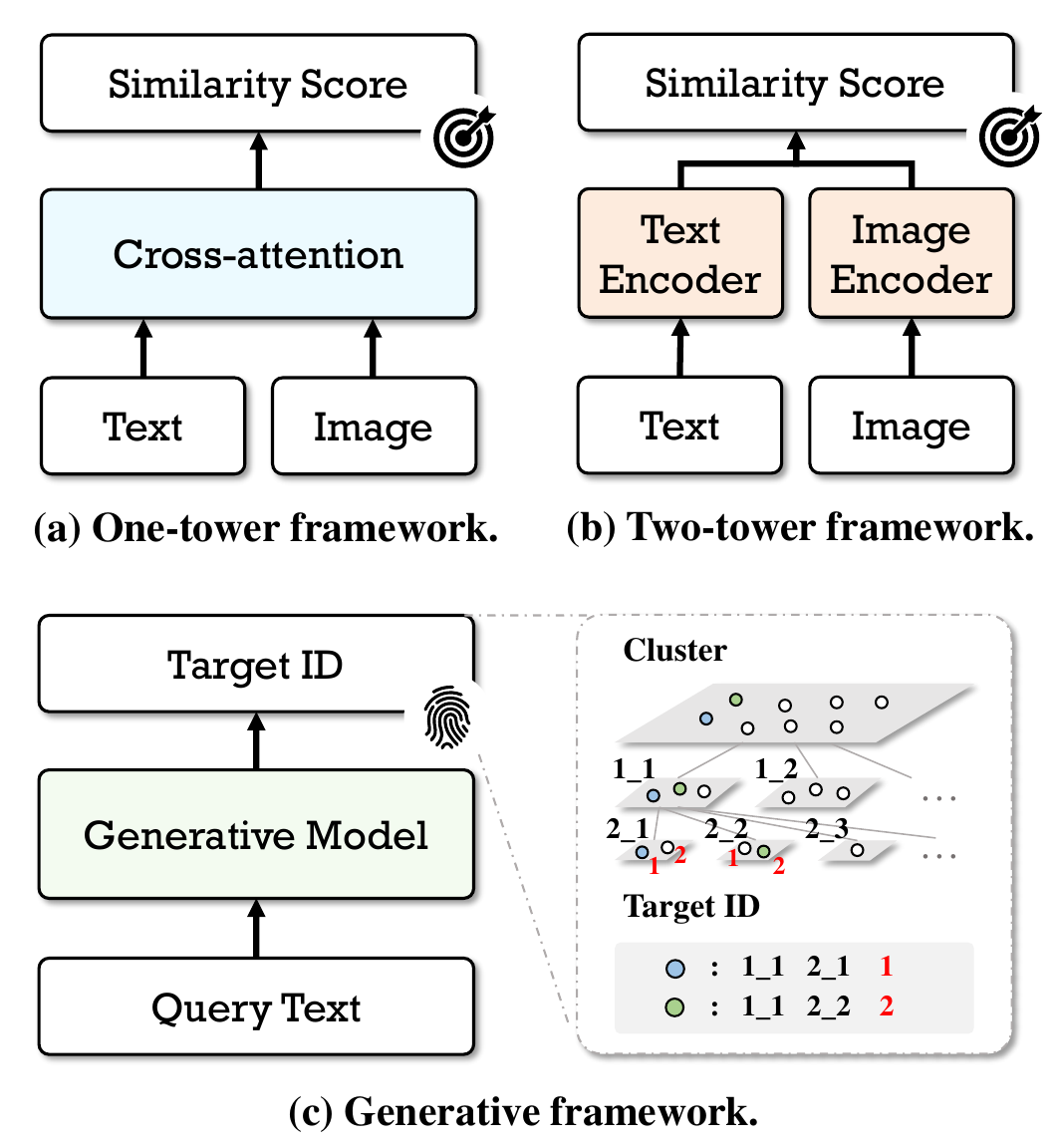}
\caption{Illustrations of existing paradigms for cross-modal retrieval. Both one-tower and two-tower frameworks perform retrieval based on certain similarity-based metrics, while the generative framework directly generates the target ID as retrieval result. Existing generative methods typically use hierarchical clustering for identifier construction.}
\label{fig:fig_1}
\end{figure}

\section{Introduction}
\label{sec:intro}

Cross-modal retrieval, as a fundamental task in multimedia research, has garnered substantial attention over the past decade~\cite{qu2021dynamic, li2023your, faghri2017vse++, chen2021learning}. This task aims to retrieve the most relevant target from one modality (\eg images) given a query from another modality (\eg text). Despite considerable advancements in recent years, cross-modal retrieval still faces challenges in achieving both accuracy and efficiency, primarily due to inherent modality heterogeneity and retrieval latency within the existing methodological paradigm.

The classical paradigm for cross-modal retrieval typically involves two major steps: first, it maps images and texts into a shared representation space, and then calculates similarity score between the two modalities to obtain the ranking list. Following this paradigm, existing solutions can be further categorized into two types of frameworks: \textbf{Two-Tower}~\cite{faghri2017vse++, chen2021learning, bin2023unifying, radford2021learning} and \textbf{One-Tower}~\cite{qu2021dynamic, lee2018stacked, zhang2022negative}, as illustrated in Figure~\ref{fig:fig_1}. Specifically, the Two-Tower independently encodes visual and textual data into a common feature space, enabling efficient retrieval through pre-embedding. However, its effectiveness is constrained by insufficient inter-modal interaction.
In contrast, the One-Tower framework leverages cross-attention mechanism to explore rich inter-modal interactions, substantially enhancing retrieval accuracy while incurring significant computational costs. Although both frameworks have demonstrated effectiveness, they still heavily rely on large-scale indices for similarity calculation, and suffer from quadratically increasing computational costs as the target set scales up, which severely limits their scalability in real-world applications.
Fortunately, the emergence of generative models~\cite{radford2018gpt1, brown2020gpt3, touvron2023llama, yao2024minicpm, radford2019gpt2} has introduced a novel retrieval paradigm known as generative retrieval, which excels in both retrieval efficiency and effectiveness through the utilization of large language models. This paradigm offers a promising alternative to traditional dense retrieval methods~\cite{li2024generative, li2024revolutionizing, zhang2023irgen}. 

Generative cross-modal retrieval, which assigns each target (\eg images) a unique identifier and trains a generative model to generate the corresponding identifier in response to an input query (\eg text) without explicit indexing. Unlike classical paradigms, where retrieval latency increases due to the growing cost of similarity computation as the target set scales up, the generative retrieval maintains consistent retrieval efficiency regardless of the target set size~\cite{li2024generative}.
Figure~\ref{fig:fig_1} illustrates this paradigmatic shift, where the classical paradigms match queries to targets based on certain similarity-based metrics, while the generative approach leverages multi-modal large language models (MLLMs)~\cite{yao2024minicpm, liu2024visual, alayrac2022flamingo} to retrieve targets by directly generating relevant identifiers. 

Despite the promising potential of generative cross-modal retrieval, current methods~\cite{li2023your, li2024revolutionizing} still face several limitations:
1) \textbf{Semantic Insufficiency in Identifier Construction.} Effective generative retrieval heavily depends on well-defined target identifiers, as generative models cannot directly analyze target content and instead rely entirely on identifiers as proxies. However, as shown in Figure~\ref{fig:fig_1} (c), existing identifiers predominantly adopt numeric-based formats, inherently restricting the capability of generative models optimized for natural language comprehension and generation. Consequently, these numeric-based identifiers fail to fully exploit the linguistic strengths of generative models, leading to suboptimal generative retrieval performance.
2) \textbf{Neglect of Fine-grained Semantics in Generative Retrieval.} Typically, similar retrieval targets tend to share consistent identifier prefixes, making the identifier suffixes crucial for distinguishing among targets. However, current approaches construct these suffixes without incorporating meaningful semantic information. For example, as illustrated in Figure~\ref{fig:fig_1} (c), the construction of hierarchical clustering identifiers terminates once it shrinks below a predefined threshold. Subsequently, identifier suffixes are randomly assigned to items within these smallest clusters, disregarding nuanced semantic distinctions. This oversight diminishes retrieval effectiveness in scenarios requiring fine-grained discrimination.

To address these limitations, we propose a novel and unified \textbf{Sem}antic-enhanced generative \textbf{C}ross-m\textbf{O}dal \textbf{RE}trieval framework (\textbf{SemCORE}), designed to unleash the semantic understanding capabilities in generative cross-modal retrieval. Specifically, we first introduce a \textbf{Structured natural language IDentifier (SID)}, comprising a Global ID and a Lexical ID.
The global ID is constructed using traditional clustering algorithms to achieve initial macroscopic retrieval, while the lexical ID employs keyword extraction techniques~\cite{grootendorst2022bertopic, grootendorst2020keybert} to facilitate more effective target localization by activating the inherent semantic understanding capability of generative models. 
Moreover, to further strengthen fine-grained semantic discrimination capability, we innovatively reformulate the final random suffixes generation into a targeted \textbf{Generative Semantic Verification (GSV)} strategy. This strategy performs nuanced distinctions at a highly detailed semantic level, thereby ensuring precise and fine-grained generative retrieval outcomes.
Additionally, our framework is the first to simultaneously consider both text-to-image and image-to-text retrieval tasks within generative cross-modal retrieval.
Compared to previous approaches, the SemCORE framework better activates the potential of generative models through richer semantic identifier constructions and fine-grained generative semantic verification strategies, ultimately enabling superior generative cross-modal retrieval.
Benefiting from this unified semantic enhancement design, SemCORE significantly outperforms existing generative cross-modal retrieval methods, achieving an average improvement of 8.65 points in Recall@1 across both datasets for text-to-image retrieval. Its performance also matches or even surpasses traditional methods for image-to-text retrieval.
The key contributions in this work are as follows:

\begin{itemize}[leftmargin=*]
    \item We introduce \textbf{SemCORE}, a novel unified semantic-enhanced generative retrieval framework that achieves comprehensive semantic enrichment throughout the retrieval process.
    \item We design a structured natural language identifier (SID) and introduce a generative semantic verification (GSV) strategy to improve semantic comprehension and fine-grained discrimination capabilities in generative cross-modal retrieval.
    \item To the best of our knowledge, SemCORE is the first framework to simultaneously consider both text-to-image and image-to-text retrieval tasks within generative cross-modal retrieval. 
    \item Extensive experiments demonstrate the superiority of our framework and underscore the importance of semantic understanding capabilities in generative cross-modal retrieval.
\end{itemize}

\section{Related Works}

\begin{figure*}
\centering
\includegraphics[width=0.98\textwidth]{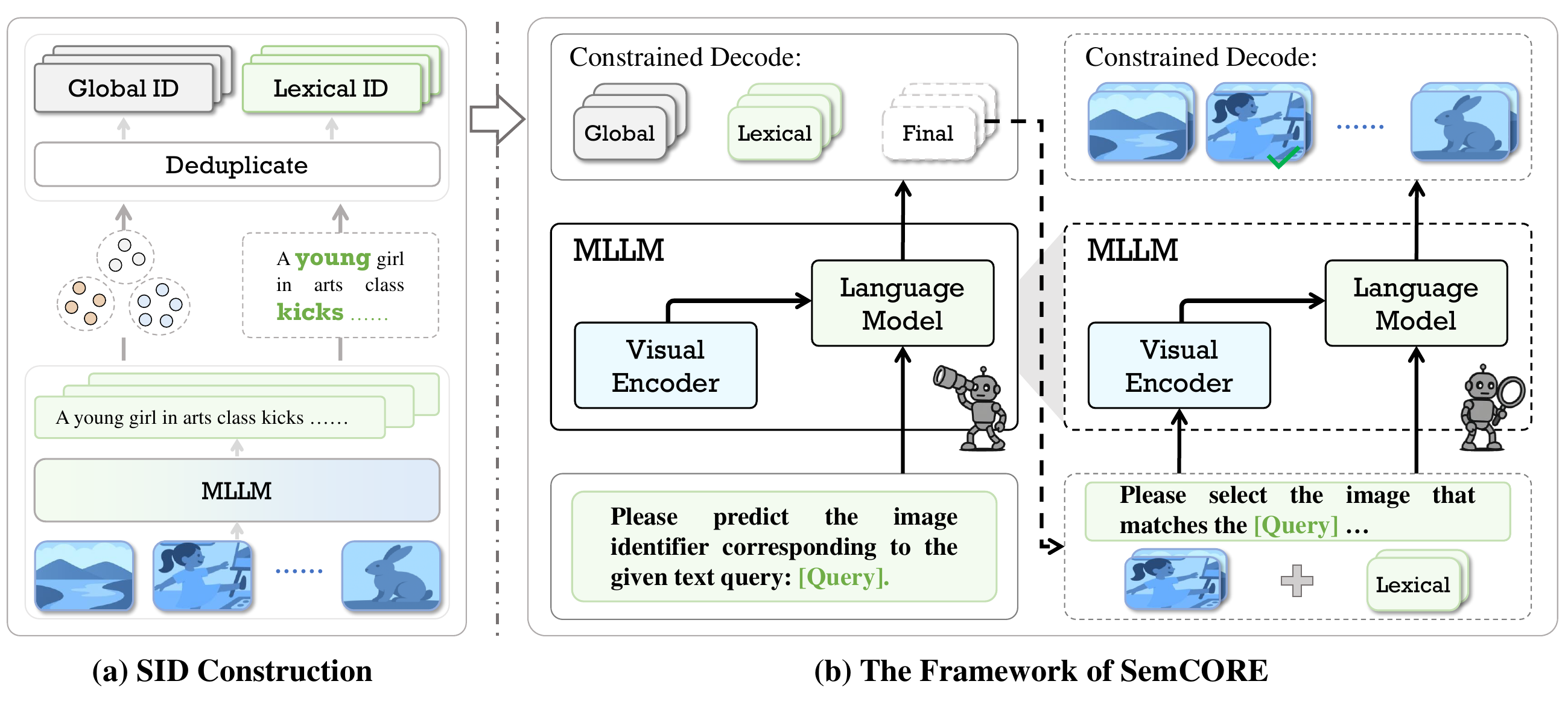}
\vspace{-0.1in}
\caption{An overview of the proposed SemCORE framework (illustrating the text-to-image retrieval process, with the image-to-text process being analogous). Specifically, we integrate a clustering algorithm with keyword extraction techniques to construct Structured natural language IDentifiers (SID), as illustrated in (a). Furthermore, we introduce a Generative Semantic Verification (GSV) strategy for nuanced semantic discrimination, as indicated by the dotted line in (b).}
\label{fig:fig_2}
\end{figure*}

\subsection{Cross-modal Retrieval}

Cross-modal retrieval can be divided into two categories according to the modality interactions, \ie, two-tower~\cite{frome2013devise, faghri2017vse++, chen2021learning, li2023your, li2022multi} and one-tower framework~\cite{lee2018stacked, qu2021dynamic, li2022blip, zhang2022negative}.
\textbf{Two-tower framework} independently projects image and text into a common feature space, where semantic similarities are calculated using cosine similarity. This approach enables efficient inference through pre-embedding, making it suitable for real-world scenarios. Frome \etal~\cite{frome2013devise} pioneered this approach by combining CNN and Skip-Gram~\cite{mikolov2013efficient} architectures for visual and language feature extraction, respectively. Faghri \etal~\cite{faghri2017vse++} proposed the VSE++ method, which integrated online hard negative mining strategy into the triplet loss function. Recently, Chen \etal~\cite{chen2021learning} introduced a Generalized Pooling Operator that optimizes the pooling strategy by learning different weights for each ranking dimension.
\textbf{One-tower framework} leverages cross-attention mechanisms to capture fine-grained interactions between visual and textual modalities. For example, the pioneering work SCAN~\cite{lee2018stacked} was proposed by Lee \etal, which measures the image-text similarity by selectively cross aggregating regions and words. Later, Qu \etal~\cite{qu2021dynamic} developed a dynamic routing mechanism that dynamically selects embedding paths based on input and implemented four specialized interaction cells for both intra-modal and inter-modal interaction. While one-tower approaches demonstrate superior performance, their reliance on cross-attention mechanisms requires extensive computation across all visual and textual data during inference, resulting in lower efficiency.


\subsection{Generative Retrieval}
Generative retrieval~\cite{tay2022transformer, wang2022neural, zhang2024generative} is an emerging paradigm in text retrieval that directly generates a target identifier as retrieval target. 
DSI~\cite{tay2022transformer} and NCI~\cite{wang2022neural}, as representative generative retrieval methods, utilize the hierarchical K-Means algorithm to construct identifiers. This methodology effectively incorporates semantic information of targets into the decoding process, thereby enhancing the modeling of the correspondence between queries and target identifiers.
In recent years, the generative retrieval paradigm has been extended to cross-modal retrieval tasks. Li \etal~\cite{li2024generative} first proposed the generative cross-modal retrieval paradigm. It explores various identifier types for images, such as Numeric ID, String ID, and Structured ID. To address the limitation of insufficient visual information in previous identifiers, Li \etal~\cite{li2024revolutionizing} subsequently proposed a novel autoregressive voken generation method, AVG. This method utilizes image tokenization to create more informative identifiers and reformulates the text-to-image retrieval task as a token-to-voken generation problem. Although significant advancements have been achieved with the aforementioned methods, generative cross-modal retrieval still suffers from certain limitations, particularly in terms of identifier construction (which remains predominantly numeric-based representation) and generative strategy (which ignores fine-grained semantics).
\section{Method}

In this section, we present a comprehensive overview of our proposed SemCORE framework, as illustrated in Figure~\ref{fig:fig_2}, which aims to unleash the potential semantic understanding capability of MLLM for advanced generative cross-modal retrieval. Specifically, we introduce two key innovations: the \textbf{Structured natural language IDentifier (SID)} and \textbf{Generative Semantic Verification (GSV)} strategy. The former integrates macroscopic and detailed semantics by combining clustering algorithms with keyword extraction techniques, while the latter leverages the inherent fine-grained semantic understanding capabilities of MLLMs for precise semantic discrimination. We first formalize the task definition and then elaborate on both components in detail.

\subsection{Preliminary}
\label{sec:pre}

In generative cross-modal retrieval, each retrieval target is assigned a unique identifier. The fundamental generative strategy involves processing a query to generate a ranked list of target identifiers, ordered by their generation probability in descending order. To ensure the validity of the generated identifier, we employ the constrained beam search algorithm~\cite{wang2022neural, li2024generative, zhang2024generative, tay2022transformer} during inference. The decoded identifiers are subsequently mapped back to their corresponding targets, yielding the final ranked retrieval results.

Formally, using text-to-image retrieval as an example, let $V$ denote the image corpus. For each image $v\in V$, its target identifier is represented as $id^v\ =\ [id_1^v, ..., id_L^v]$, where $L$ represents the length. The generative model $D$ sequentially generates the corresponding target identifier for any given text query $t$. At each step, the model generates the next identifier token conditioned on the previously generated tokens and the original text query. The relevance between text query $t$ and each image is calculated as follows:
\begin{equation}
\label{equ:relevance}
R(t, v)=D_\theta(id^v|t) = \prod^L_{i=1}D_\theta(id_i^v|id_{<i}^v,t),
\end{equation}
where $id_i^v$ is the $i^{th}$ target identifier token and $id_{<i}^v$ represents previously generated tokens, denoted as $id_{<i}^v = [id_1^v, ..., id_{i-1}^v]$. Finally, images are ranked according to their relevance scores $R(t, v)$, with the top-ranked image returned as the retrieval result.

\subsection{Structured Natural Language Identifier}
Identifier quality is crucial in generative retrieval. Existing generative cross-modal retrieval methods typically construct identifiers either through hierarchical clustering algorithm~\cite{li2024generative} or RQ-VAE~\cite{lee2022autoregressive, li2024revolutionizing} encoding technique. However, these methods depend on numerical representations, inherently restricting the capabilities of generative models optimized for natural language comprehension and generation.
Conversely, directly using keywords as identifiers, as done in some generative document retrieval methods~\cite{zhang2024generative, li2023multiview}, may lead to a loss of semantic information. To address these limitations, we propose integrating a clustering algorithm with a keyword extraction technique to construct structured natural language identifiers. As shown in Figure~\ref{fig:fig_2}, these identifiers preserve complete semantic information as well as natural language representations, effectively activating the semantic understanding capabilities of pre-trained MLLMs.

\subsubsection{Global ID} 
Intuitively, the first token of an identifier plays a critical role, as its accuracy directly influences overall generation validity. However, a single keyword often inadequately represents the complete semantics of the target object. Thus, we propose to construct the Global ID through a clustering algorithm to capture macroscopic semantic information. Specifically, we apply the K-Means clustering~\cite{hartigan1979algorithm} algorithm to the embeddings of all the target items. For images, to prevent information leakage and to establish a foundation for constructing lexical ID, we utilize a MLLM to automatically generate image captions, which are used to construct the identifiers.
The clustering result serves as the first token of the structured natural language identifier.

\subsubsection{Lexical ID} 
Subsequently, to enable more effective initial target localization, we further construct Lexical ID using the KeyBERT~\cite{grootendorst2020keybert} algorithm. This process identifies the most semantically relevant sub-phrases from target captions based on BERT embeddings and cosine similarity. Specifically, we first extract embeddings using BERT~\cite{devlin2018bert} to obtain target-level representations, then extract embeddings of n-gram sub-phrases and calculate their cosine similarity with the overall target embedding. The sub-phrases with the highest semantic similarity are selected as lexical IDs.
Additionally, to avoid semantic redundancy between lexical ID and global ID, we implement a semantic deduplication mechanism. During global ID construction, each cluster is treated as a single unit, and TF-IDF~\cite{salton1988term} is applied to the aggregated captions within each cluster to filter keywords with high weights. The TF-IDF is as follows:
\begin{equation}
    W_{x,cluster}=\|tf_{x,cluster}\|\times log(1+\frac{A}{f_x}),
\end{equation}
where $tf_{x,cluster}$ is the frequency of word $x$ in the cluster, $f_x$ is the frequency of word $x$ across all clusters, and $A$ is the average number of words per cluster. Following that, we remove candidate keywords that overlap with high-frequency cluster keywords during lexical ID construction, enhancing semantic diversity and distinction.
Finally, the structured natural language identifier is formed by concatenating global ID and lexical ID, integrating both macroscopic and detailed semantics, and has the inherent advantage of effectively activating the semantic understanding capabilities of pre-trained MLLMs. Experiment details and hyperparameter configurations will be presented in the supplementary materials.

\subsection{Generative Semantic Verification}
\label{sec:gsv}
In large-scale retrieval scenarios, it is common to have semantically close items, which makes it difficult to avoid identifier collisions~\cite{rajput2023recommender}. Previous generative cross-modal retrieval methods typically rely on maintaining an ID table to detect conflicts. When duplicate identifiers occur, the ID table records their occurrences and appends this value to the identifier as a final token to ensure uniqueness. A representative example is the hierarchical clustering-based identifier construction method, which terminates once the number of samples falls below a predefined threshold. Subsequently, identifier suffixes are randomly assigned to items within these smallest clusters, disregarding nuanced semantic differences between individual samples. To address this limitation, we propose a generative semantic verification strategy that leverages the inherent fine-grained semantic understanding capabilities of MLLMs for precise discrimination among semantically similar candidates. 

The verification process involves the following key steps: 1) Candidate Aggregation. After initial target localization using structured natural language identifier, we obtain a set of candidate targets $\mathcal{C} = \{c_1, c_2, ..., c_k\}$. These candidates typically share similar global IDs and potentially similar lexical IDs. 2) Verification Prompt Construction. To guide the model in analyzing the semantic relevance between the query and each candidate, we design specialized verification prompts. The prompt template (text-to-image direction) is structured as follows:
% ...
\begin{Verbatim}[commandchars=\\\{\}]
\textbf{USER}:
Image:
[\textcolor{blue}{Image from Candidate}]; [\textcolor{blue}{Lexical ID of Image}].
...
[\textcolor{blue}{Image from Candidate}]; [\textcolor{blue}{Lexical ID of Image}].
Sentence: [\textcolor{blue}{Text Query}].
Please select the image that best matches the last se-
ntence based on the image with its keywords. 
\textbf{ASSISTANT:}
[\textcolor{blue}{Target Image}]
\end{Verbatim}

\noindent where the text in the middle brackets refers to specific objects, not specific text. The prompt explicitly instructs the model to analyze semantic relationship between the query and each candidate, considering fine-grained intra-modal features and cross-modal semantic relationships (while embedding generated lexical ID to further activate the semantic understanding capability of MLLMs). 
3) Fine-grained Semantic Discrimination. MLLMs process the verification prompts and perform semantic discrimination to determine the most relevant candidate. Experimental details and hyperparameter configurations will be presented in the supplementary materials.

\subsection{Training and Inference}
\textbf{Training.} Considering factors including convenience and model capabilities, we have chosen InternVL2.5~\cite{chen2024expanding} as the backbone for our framework. To implement the structured natural language identifier generation, we leverage <query, identifier> pairs to train the model. The training loss function is as follows.
\begin{equation}
    \mathbf{\mathcal{L}}=-\sum_{i=1}^L log\ D_{\theta}(id^{v}_i|id_{<i}^v,query).
\end{equation}
For the training of generative semantic verification strategy, we employ the autoregressive language modeling objective, specifically next token prediction training loss.

\begin{algorithm}
    \caption{Constrained Decoding Algorithm.}
    \small
    \begin{algorithmic}[1]
        \STATE $\mathcal{T}\gets Trie\ Structure$
        \FOR{$v\in V$}
        \STATE $\mathcal{T}.add(SID(v))$
        \ENDFOR
        \STATE $B\gets Beam\ Size$
        \FOR{$1 \leq b \leq B$}
        \STATE $X_{b,0}\gets [<BOS>]$
        \STATE $Z_{b,0}\gets \mathcal{T}.get\_valid\_token(\emptyset)$
        \ENDFOR
        \FOR{$1\leq t \leq L$}
        \STATE $\forall 1\leq b \leq B,\ \chi_{b,t}\gets X_{b,t-1}\times Z_{b,t-1}$
        \STATE $X_{1,t},...,X_{B,t}\gets \mathop{\arg\max}\limits_{X_{b,t}\in \chi_{b,t}} \sum\limits_{b=1}\limits^{B} log\ p(X_{b,t}|X_{b,<t};Q;\theta)$
        \STATE $\forall 1 \leq b \leq B,\ Z_{b, t}\gets \mathcal{T}.get\_valid\_token(X_{b,t})$
        \ENDFOR
        \RETURN $Top\{X_{b,L}\}_{b=1}^{B}$
    \end{algorithmic}
    \label{alg_1}
\end{algorithm}

\noindent \textbf{Inference.} The inference process follows the procedure outlined in section~\ref{sec:gsv}. Here, we detail the constrained decoding algorithm~\cite{wang2022neural, li2024generative, zhang2024generative, tay2022transformer, zeng2024planning}, which restricts the generation space to targets contained in the test set, ensuring the validity and interpretability of generated results. This approach constructs a Trie structure to store all legal target identifiers, enabling rapid matching and filtering during generation. Given a prefix sequence during generation, the Trie dynamically suggests all possible legal next tokens, guiding the model to produce identifiers conforming to constraints. This ensures every generated identifier corresponds to an entity in the corpus. The overall generation process is shown in Algorithm~\ref{alg_1}, where beam size is denoted as $B$ and the identifier length as $L$. For clarity, we define the following variables:
\begin{itemize}[leftmargin=*]
    \item $X_{b,t}$: the token sequence generated in the $b$-th beam up to the $t$-th generation step.
    \item $Z_{b,t}$: the set of valid next tokens constrained by the Trie at the $t$-th step for $b$-th beam.
\end{itemize}
This constrained decoding strategy significantly enhances the accuracy of generated results.

\section{Experiments}





We conduct comprehensive experiments to evaluate the effectiveness of our proposed SemCORE framework and answer the following research questions:
\begin{itemize}[leftmargin=*]
    \item \textbf{RQ1:} How does our proposed SemCORE compare to existing generative cross-modal retrieval methods?
    \item \textbf{RQ2:} What are the contributions of individual components within SemCORE?
    \item \textbf{RQ3:} How does the performance of SemCORE vary under different configurations, including cluster size, identifier length, and model scale?
\end{itemize}

\subsection{Experimental Setup}
We first describe our experimental configuration, including datasets and the baselines compared.

\subsubsection{Datasets}
We evaluate our framework on two widely-used datasets, Flickr30K~\cite{young2014image} and MS-COCO~\cite{lin2014microsoft}. Flickr30K comprises 31,783 images, each of which is associated with 5 different sentences. According to the previous settings in~\cite{lee2018stacked, liu2019focus}, we split this dataset into 29,783 training images, 1,000 validation images, and 1,000 testing images. MS-COCO dataset comprises 123,287 images, each of which is also annotated with five different sentences. To ensure a fair comparison with previous works, we used the Karpathy split~\cite{karpathy2015deep} to divide the dataset into 113,287 training images, 5,000 validation images, and 5,000 testing images. For evaluation on Flickr30K and MS-COCO datasets, we follow previous works~\cite{lee2018stacked, chen2021learning} and adopt R@K (Recall@K, K=1,5) and rSum as evaluation metrics for retrieval results, with a focus on fine-grained retrieval. R@K measures the percentage of queries for which the ground-truth hit in its top-K ranking list and rSum means the sum of all R@K. Higher R@K and rSum values indicate superior performance.

\subsubsection{Baselines}
As generative cross-modal retrieval is a relatively new research area primarily focused on text-to-image tasks~\cite{li2024generative, li2024revolutionizing}, we select state-of-the-art methods from both generative and traditional paradigms for comprehensive comparison.

\noindent \textbf{Text-to-Image Retrieval:} Following~\cite{li2024revolutionizing}, we select the following generative cross-modal retrieval methods as the baselines:
\begin{itemize}[leftmargin=*]
    \item \textbf{GRACE~\cite{li2024generative}:} A pioneering generative retrieval framework utilizing various identifier construction types. Consistent with~\cite{li2024revolutionizing}, we also exclude the ``Atomic ID'' variant as it falls outside the generative retrieval paradigm.
    \item \textbf{IRGen~\cite{zhang2023irgen}:} Originally proposed for image-to-image retrieval, this method has been adapted by~\cite{li2024revolutionizing} for text-to-image retrieval by substituting its original image input with text input.
    \item \textbf{AVG~\cite{li2024revolutionizing}:} This method discretizes images into a sequence of vokens as target identifiers, formulating text-to-image retrieval as token-to-voken generation.
\end{itemize}

\noindent \textbf{Image-to-Text Retrieval:} We reproduce the GRACE~\cite{li2024generative} method and select two representative approaches from the traditional paradigm for comparison:
\begin{itemize}[leftmargin=*]
    \item \textbf{One-tower: LeaPRR~\cite{qu2023learnable},} a novel learnable pillar-based re-ranking method for image-text retrieval that leverages multi-modal neighbor relations to improve retrieval performance.
    \item \textbf{Two-tower: FNE~\cite{li2023your},} this method introduces a novel training strategy to alleviate the false negative issue in cross-modal retrieval. It employs BERT~\cite{devlin2018bert} and ViT~\cite{dosovitskiy2020image} as the encoders for textual and visual modalities, making it a representative method among two-tower approaches.
\end{itemize}
AVG~\cite{li2024revolutionizing}, the best-performing generative retrieval method, is omitted from image-to-text experiments due to its reliance on LLMs instead of MLLMs, rendering it fundamentally unsuitable for image-to-text retrieval.
The implementation details will be presented in the supplementary materials. 

\begin{table}[t]
\centering
\caption{Performance comparison between SemCORE and other generative cross-modal retrieval methods on text-to-image retrieval. The best results are marked in bold. 
}
\label{tab:tab1}
\begin{tabular}{cccc}
\toprule
{Methods} & {R@1} & {R@5} & {rSum} \\
\midrule
\multicolumn{4}{c}{{Flickr30K}} \\
\midrule
GRACE~\cite{li2024generative} (Numeric ID)    &22.5 &28.9 &51.4  \\
GRACE~\cite{li2024generative} (String ID)     &30.5 &39.0 &69.5  \\
GRACE~\cite{li2024generative} (Semantic ID)   &22.9 &34.9 &57.8  \\
GRACE~\cite{li2024generative} (Structured ID) &37.4 &59.5 &96.9  \\
IRGen~\cite{zhang2023irgen}                   &49.0 &68.9 &117.9 \\
AVG~\cite{li2024revolutionizing}              &62.8 &\textbf{85.4} &148.2 \\
\textbf{SemCORE (Ours)} &\textbf{69.0} &83.0 &\textbf{152.0} \\
\midrule
\multicolumn{4}{c}{{MS-COCO (5k)}}   \\
\midrule
GRACE~\cite{li2024generative} (Numeric ID)    &0.03 &0.14 &0.17 \\
GRACE~\cite{li2024generative} (String ID)     &0.12 &0.37 &0.49 \\
GRACE~\cite{li2024generative} (Semantic ID)     &13.3 &30.4 &43.7 \\
GRACE~\cite{li2024generative} (Structured ID) &16.7 &39.2 &55.9 \\
IRGen~\cite{zhang2023irgen}                   &29.6 &50.7 &80.3 \\
AVG~\cite{li2024revolutionizing}              &31.3 &\textbf{58.0} &89.3 \\
\textbf{SemCORE (Ours)} &\textbf{42.4} &57.5 &\textbf{99.9} \\                      
\bottomrule
\end{tabular}
\end{table}

\begin{table}[t]
\setlength{\tabcolsep}{5pt}
\centering
\caption{Performance comparison of image-to-text retrieval on Flickr30K dataset against selected baselines. For the generative paradigm, the best results are highlighted in bold. $*$ denotes the best result from the GRACE reproduction.
}
\label{tab:tab2}
\begin{tabular}{llccc}
\toprule
\multirow{2}{*}{Paradigm} & \multirow{2}{*}{Methods} & \multicolumn{3}{c}{Image-to-Text} \\ 
&& R@1 & R@5 & rSum \\ 
\midrule
One-tower  &LeaPRR~\cite{qu2023learnable}  &86.2 &96.6 &182.8 \\ 
Two-tower  &FNE~\cite{li2023your}          &85.4 &98.0 &183.5 \\ 
\midrule
\multirow{2}{*}{Generative} 
&GRACE$^*$~\cite{li2024generative}         &23.6 &35.8 &59.4 \\ 
&\textbf{SemCORE (Ours)}                   &\textbf{88.8} &\textbf{96.1} &\textbf{184.9} \\ 
\bottomrule
\end{tabular}
\end{table}

\subsection{Performance Comparison (RQ1)}
To evaluate the effectiveness of our proposed SemCORE framweork (RQ1), we conduct extensive experiments on Flickr30K and MS-COCO (5k) to compare the performance of SemCORE with existing generative cross-modal retrieval methods.

\subsubsection{Performance \wrt Text-to-Image Retrieval.}
The results of text-to-image retrieval are presented in Table~\ref{tab:tab1}. We can observe that SemCORE significantly outperforms all generative baselines on the most important metric, Recall@1, achieving an average improvement of 8.65 points across both datasets. Notably, on the more challenging MS-COCO dataset with its larger test set, our proposed SemCORE achieves an even greater improvement (11.1 vs 6.2 on Flickr30K). Furthermore, SemCORE also achieves the best performance in terms of rSum. These findings collectively demonstrate the superiority of our proposed SemCORE framework. However, it should be noted that our method performs slightly worse than AVG on the R@5 metric. This limitation can be attributed to the fact that the SIDs we constructed are based on natural language expressions, inevitably suffering from issues such as synonym ambiguity that can confuse generative retrieval. Nevertheless, this slight reduction in R@5 represents a reasonable trade-off for the significant improvement in R@1, which is the most critical indicator of retrieval precision.
Additionally, we can observe that GRACE underperforms compared to IRGen and AVG. This is because GRACE does not explicitly incorporate target semantic information when constructing the target identifier. While the latter two methods construct target identifiers using learned image tokenizers, thereby establishing stronger connections with target images, they still fall short of our approach. The reason lies in the fact that, although previous methods have considered target semantics during identifier construction, they still fall short in terms of identifier representation and fine-grained semantic modeling. Our framework addresses these limitations, more effectively activating the semantic understanding capabilities of MLLMs pre-trained on natural language. 

\subsubsection{Performance \wrt Image-to-Text Retrieval.}
To comprehensively validate the superiority of our framework, we also evaluate SemCORE on image-to-text retrieval tasks. The performance comparison is presented in Table~\ref{tab:tab2}. It can be observed that although generative cross-modal retrieval emerging as a new paradigm that leverages the powerful capabilities of multi-modal large language models, existing methods (\eg GRACE) still demonstrate performance gaps compared to traditional one-tower or two-tower approaches. In contrast, to the best of our knowledge, our proposed SemCORE is the first framework to bridge this gap. It achieves retrieval performance on par with, or even surpassing (Recall@1), traditional methods in cross-modal retrieval. This further illustrates the effectiveness of our SemCORE framework.

\begin{table}[t]
\centering
\caption{Ablation study on key components of SemCORE.}
\label{tab:tab3}
\begin{tabular}{lccc}
\toprule
\multirow{2}{*}{Methods} & \multicolumn{3}{c}{Flickr30K} \\
& 
R@1 & R@5 & rSum \\
\midrule
SemCORE                   &69.0 &83.0 &152.0 \\
\midrule
w/o GCV                   &51.9 &77.9 &129.8  \\
w/o SID                   &66.1 &79.0 &145.1  \\
w/o global ID             &65.7 &78.2 &143.9  \\
w/o deduplication         &68.1 &81.1 &149.2  \\
w/o constrained decoding  &48.3 &55.7 &104.0  \\
\bottomrule
\end{tabular}
\end{table}

\subsection{Ablation Study (RQ2)}
To valuate the contribution of each component of the proposed framework (RQ2), we conduct comprehensive ablation studies on the Flickr30K dataset to evaluate the contribution of each component within our proposed framework. The experimental results are reported in Table~\ref{tab:tab3}. Specifically, we perform detailed analyses on the following variants.

1) For the core components of SemCOER, we individually validate their contributions: 
\textbf{\textit{w/o GSV}}, this setting directly removes the generative semantic verification strategy, substituting it with randomly assigned identifier suffixes. Retrieval performance decreases significantly due to the absence of fine-grained semantic information, particularly impacting the Recall@1 metric. This phenomenon underscores the importance of fine-grained semantic information and validates the effectiveness of our proposed strategy.
\textbf{\textit{w/o SID}}, this setting replaces the keywords within structured natural language identifiers with anonymous unique numbers. The performance degradation demonstrates that our proposed structured natural language identifiers effectively activate the inherent semantic understanding capabilities of MLLMs. Furthermore, we argue that the performance decline also relates to the absence of an explicit target-identifier training process in our method. Without SID, the model struggles to learn precise query-target correspondences, resulting in significantly decreased performance.

2) The contribution of SID does not solely rely on natural language representation, hence, we conduct a more detailed validation. \textbf{\textit{w/o global ID}}, this setting eliminates the global ID from SID, retaining only lexical ID for identifier construction. Experiment results indicate that explicitly incorporating macroscopic semantic information positively impacts the initial stages of generative retrieval, improving the accuracy of target localization.
\textbf{\textit{w/o deduplication}}, in this setting, semantic deduplication relative to the global ID is omitted when constructing lexical IDs. The performance decrease demonstrates that semantic deduplication enhances identifier discriminability, thereby improving retrieval outcomes.

3) \textbf{\textit{w/o constrained decoding}}, in this setting, the decoding process does not employ the Trie structure for constraints, enabling auto-regressive generation from the entire vocabulary. We can observe a significant drop in retrieval performance, which is consistent with previous work~\cite{li2024generative, wang2022neural}. Besides the uncertainty increase caused by the excessively large decoding space, another potential reason for the performance degradation is synonym substitution issues of keywords within our lexical ID. Generated identifiers may structurally resemble ground-truth identifiers, but even minor character deviations can prevent accurate mapping to the target.

\subsection{In-Depth Analysis (RQ3)}
To answer the question RQ3, we examine SemCORE under various configurations, including cluster size, identifier length, and model scale. Furthermore, we provide an in-depth analysis of the experimental results and specific cases.

\begin{table}[t]
\centering
\caption{Performance of different model backbones on the Flickr30K dataset. The best results are highlighted in bold. 
}
\label{tab:tab4}
\begin{tabular}{lcccc}
\toprule
Backbone &Finetuning &R@1 &R@5 &rsum \\
\midrule
InternVL2.5-1B &Full &69.0 &83.0 &152.0 \\
InternVL2.5-2B &Full &70.6 &82.3 &152.9 \\
InternVL2.5-4B &LoRA &74.5 &83.3 &157.8 \\
InternVL2.5-8B &LoRA &\textbf{75.8} &\textbf{84.5} &\textbf{160.3} \\
\bottomrule
\end{tabular}
\end{table}

\begin{figure}
\centering
\includegraphics[width=0.45\textwidth]{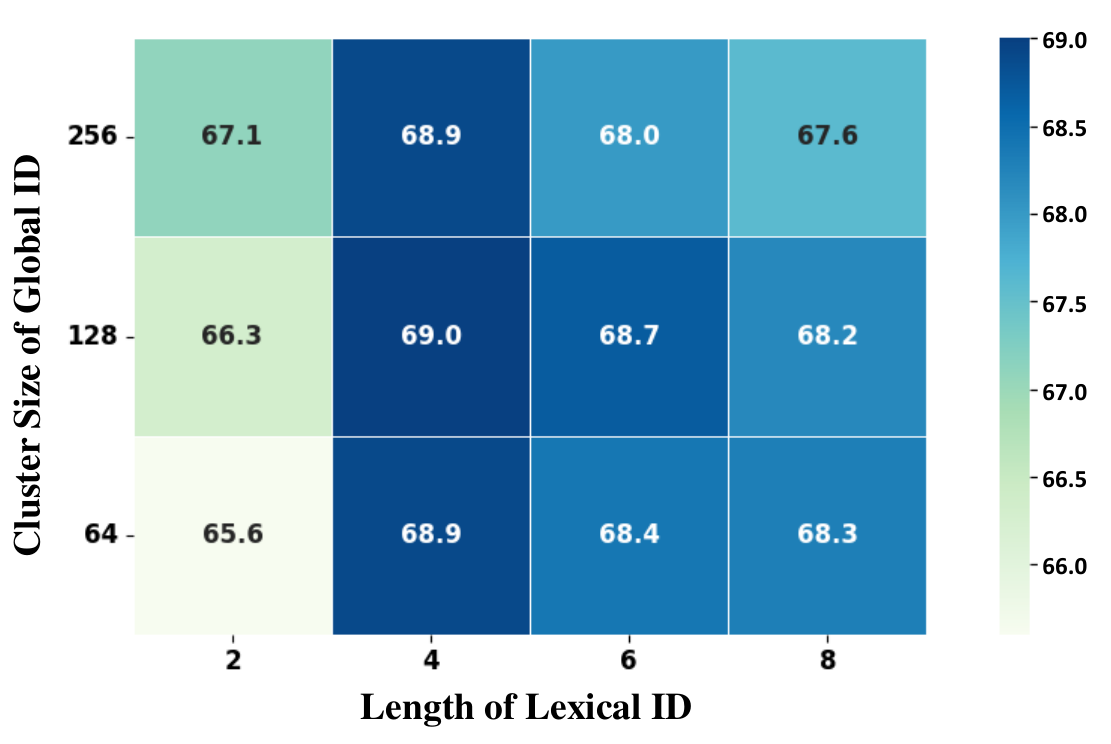}
\caption{Performance with respect to the cluster size of global ID and the length of lexical ID on Flickr30K dataset.
}
\label{fig:fig_3}
\end{figure}

\begin{figure*}[ht]
\centering
\includegraphics[width=0.95\textwidth]{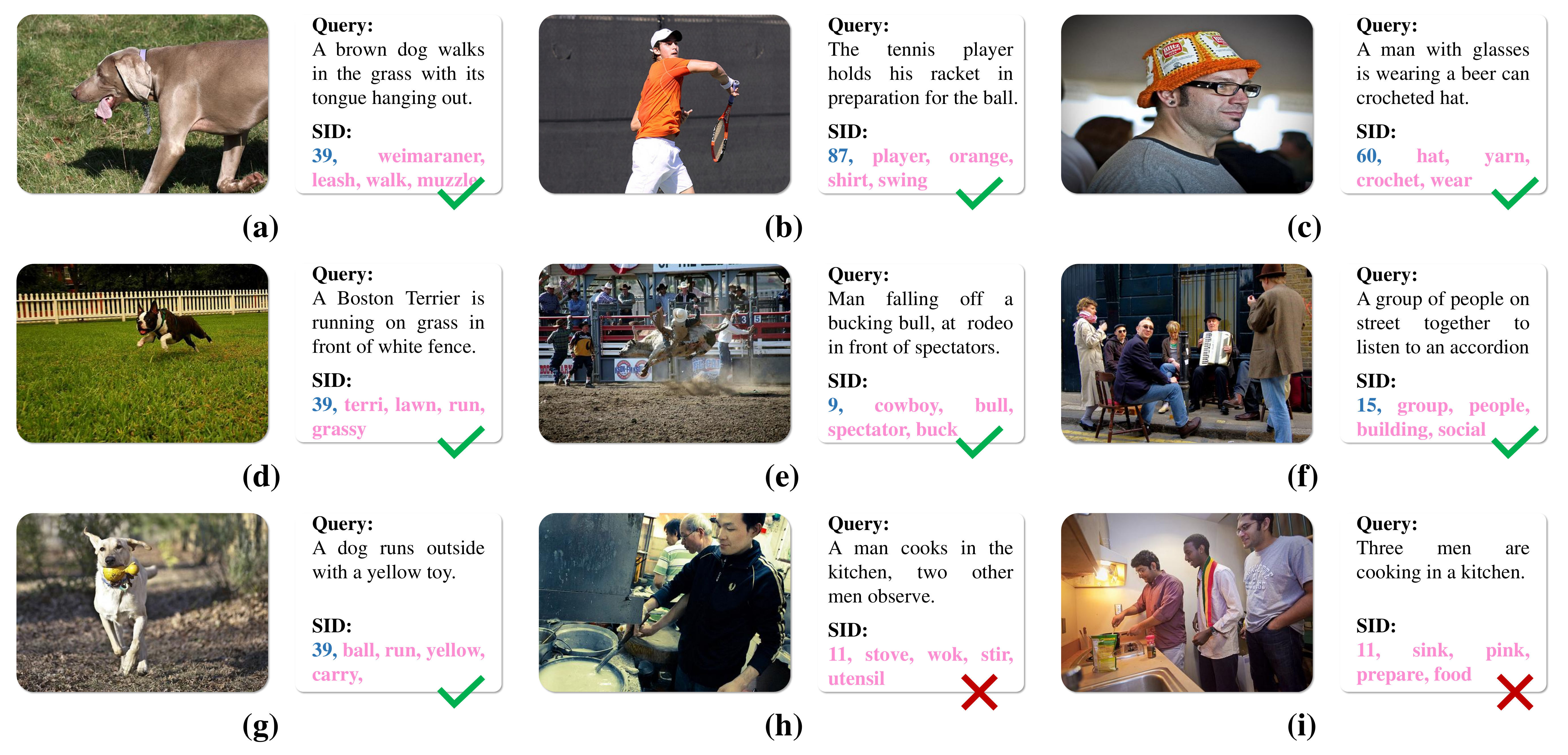}
\caption{Illustration of the structured natural language identifier (SID). The SID comprises two components: global ID and lexical ID. Global IDs are highlighted in blue, while lexical IDs are highlighted in pink. The generated lexical IDs are closely aligned with the corresponding image content. Subfigures (g) and (h) generate the SID of each other.}
\label{fig:fig_4}
\end{figure*}

\subsubsection{Analysis on Identifier Configuration.} 
Defining appropriate target identifiers is fundamental for achieving effective generative cross-modal retrieval. Our proposed structured natural language identifier consists of global ID and lexical ID. We conduct experiments on Flickr30k dataset to investigate the effects of global identifier clustering size and lexical ID length on retrieval performance, with results illustrated in Figure~\ref{fig:fig_3}. It can be observed that retrieval performance remains relatively stable across different clustering sizes, except when lexical ID length equals 2. We attribute this to the fact that shorter lexical ID length might insufficiently represent semantic information, thereby weakening semantic discrimination. Larger clustering sizes can partially compensate for this deficiency, explaining the heightened sensitivity to clustering size variations when lexical ID length is minimal. Moreover, we can also observe the optimal retrieval performance occurs with a lexical ID length of 4. We argue that overly short lengths might lead to inadequate semantic discrimination, while longer lengths, despite enhanced semantic expression, could introduce more noise and error accumulation that negatively impact performance. Once the identifier length sufficiently supports semantic discrimination (\eg length = 4), further increasing the length typically yields diminishing returns. Therefore, based on these results, the cluster size of 128 and lexical ID length of 4 are optimal configuration values.

\subsubsection{Analysis on Model Scale.}
Generally, the overall performance of MLLMs improves with increasing model size, a trend known as the scaling law~\cite{kaplan2020scaling}. To examine whether this principle applies to generative cross-modal retrieval, we experiment with various backbones, such as InternVL2.5-1B, InternVL2.5-2B, InternVL2.5-4B, and InternVL2.5-8B. Due to computing resource constraints, we adopt the LoRA~\cite{hu2022lora} technique for fine-tuning the larger InternVL2.5-4B and 8B models. The results are summarized in Table~\ref{tab:tab4}. Experimental results indicate a generally improved retrieval performance with increasing model size, suggesting that the scaling law remains applicable in generative cross-modal retrieval. This phenomenon is also consistent with the expectation that larger MLLMs typically possess stronger semantic understanding and generative capabilities. However, performance gains at the 8B scale fall below expectations, which we attribute to limitations inherent in the LoRA technique. Since LoRA updates only a small fraction of model parameters, it restricts the power of model to learn the correspondences between queries and identifiers, a critical aspect in generative retrieval tasks. Consequently, limited fine-tuning capabilities likely constrains further performance improvements.

\subsubsection{Case Study.}
To further analyze the retrieval results of our proposed framework with respect to structured natural language identifiers, we present several representative examples in Figure~\ref{fig:fig_4} (text-to-image retrieval). The constructed SID highly correlates with the image contents and effectively captures key visual elements.
Specifically, as shown in Figure~\ref{fig:fig_4} (a), (d), and (g), identical global IDs (marked in blue) indicate that the model successfully captures the macro-level semantic information, while lexical IDs (marked in pink) further refine this by highlighting details, such as specific dog breeds. This SID construction facilitates effective generative cross-modal retrieval.
Notably, we also observe some failure cases, as illustrated in Figure~\ref{fig:fig_4} (h) and (i), these two instances generate identifiers corresponding to each other. However, even in such incorrect generations, the textual queries still maintain semantic relevance to the image linked to the incorrect identifiers. This phenomenon resembles the "false negative" issue raised in~\cite{li2023your}, revealing inherent diversity and flexibility within the semantic matching between images and texts. In practical applications, the same image may reasonably correspond to multiple textual descriptions, and vice versa. Consequently, these seemingly incorrect matches may stem more from limitations in dataset annotation standards rather than semantic-level failures of the model. In conclusion, these observations validate the robustness of our method when dealing with semantic diversity and further demonstrate its capabilities in semantic understanding and matching, showcasing its potential for applications beyond traditional evaluation metrics. We also illustrate more examples in the supplementary material, \wrt the examples of image-to-text retrieval.
\section{Conclusion}
In this paper, we proposed SemCORE, a novel unified semantic-enhanced generative cross-modal retrieval framework. Specifically, we constructed a structured natural language Identifier (SID), comprising a coarse-grained global ID derived from a clustering algorithm and a fine-grained lexical ID extracted through a keyword extraction technique. This identifier effectively activates the inherent semantic understanding capabilities in generative models while maintaining global semantic information. 
Furthermore, to address the neglect of fine-grained semantic information in existing generative retrieval approaches, we introduced a generative semantic verification (GSV) strategy, which leverages MLLMs to perform fine-grained semantic discrimination. Notably, SemCORE is the first framework to simultaneously consider both text-to-image and image-to-text retrieval tasks within the generative cross-modal retrieval paradigm. Extensive experiments conducted on Flickr30K and MS-COCO demonstrate the effectiveness of our framework. 
In the future, we plan to further explore the generalization capability of generative retrieval. While current methods predominantly rely on LLMs to memorize static mappings between queries and identifiers, real-world retrieval sets are typically dynamic. Therefore, improving the generalization capacity of generative retrieval represents a critical direction toward enabling practical deployment.


\begin{acks}
We are grateful to Yongqi Li for his valuable discussions and feedback on this work.
\end{acks}

\balance
\bibliographystyle{ACM-Reference-Format}
\bibliography{acmart}

\clearpage
\appendix
\section{Appendix}

\subsection{Implementation Details}
Considering factors including convenience and model capabilities, we have chosen InternVL2.5 as the backbone for our framework. InternVL2.5 is a series of MLLMs designed for vision-language understanding tasks. The training process follows the official configuration recommended by InternVL. Specifically, we conducted model training based on the DeepSpeed distributed training framework using two NVIDIA A40 GPUs. During fine-tuning, we employed both full-parameter fine-tuning and the LoRA algorithm, with a total of 5 training epochs and a batch size of 4 per GPU. For the training of generative semantic verification (GSV) strategy, we constructed candidate sets through generative retrieval based on a model previously trained with structured natural language identifiers (SID). During SID construction, the global ID clustering algorithm was configured with a cluster size of 128. For lexical ID construction, we implemented deduplication based on the top five keywords extracted via TF-IDF. The candidate set size for the GSV strategy was fixed at 10.

\subsection{Analysis on Beam Search}
In generative cross-modal retrieval paradigm, beam search is typically employed to generate a ranked list of candidate targets. Therefore, investigating the impact of beam width on retrieval performance is important. We summarize the retrieval performance of SemCORE under different beam widths in Table~\ref{tab:tab5}. Experimental results indicate that retrieval effectiveness generally improves as the beam width increases. This improvement can primarily be attributed to the reduced likelihood of missing the correct target with a larger beam width. However, increasing the beam width also leads to efficiency loss, and the performance gain tends to plateau when the beam width exceeds 20. Consequently, considering both retrieval effectiveness and efficiency, selecting a beam width of 20 or 30 represents a more reasonable choice.


\begin{table}[th]
\centering
\caption{Performance of SemCORE versus the beam size. The best results are marked in bold.}
\label{tab:tab5}
\begin{tabular}{cccc}
\toprule
Beam Size &R@1 &R@5 &rsum \\
\midrule
10 &64.4 &76.5 &140.9 \\
20 &67.5 &80.9 &148.4 \\
30 &68.3 &82.5 &150.8 \\
40 &68.8 &82.8 &151.6 \\
50 &\textbf{69.0} &\textbf{83.0} &\textbf{152.0} \\
\bottomrule
\end{tabular}
\end{table}

\subsection{Comprehensive Case Study}
\begin{figure*}[ht]
\centering
\includegraphics[width=0.8\textwidth]{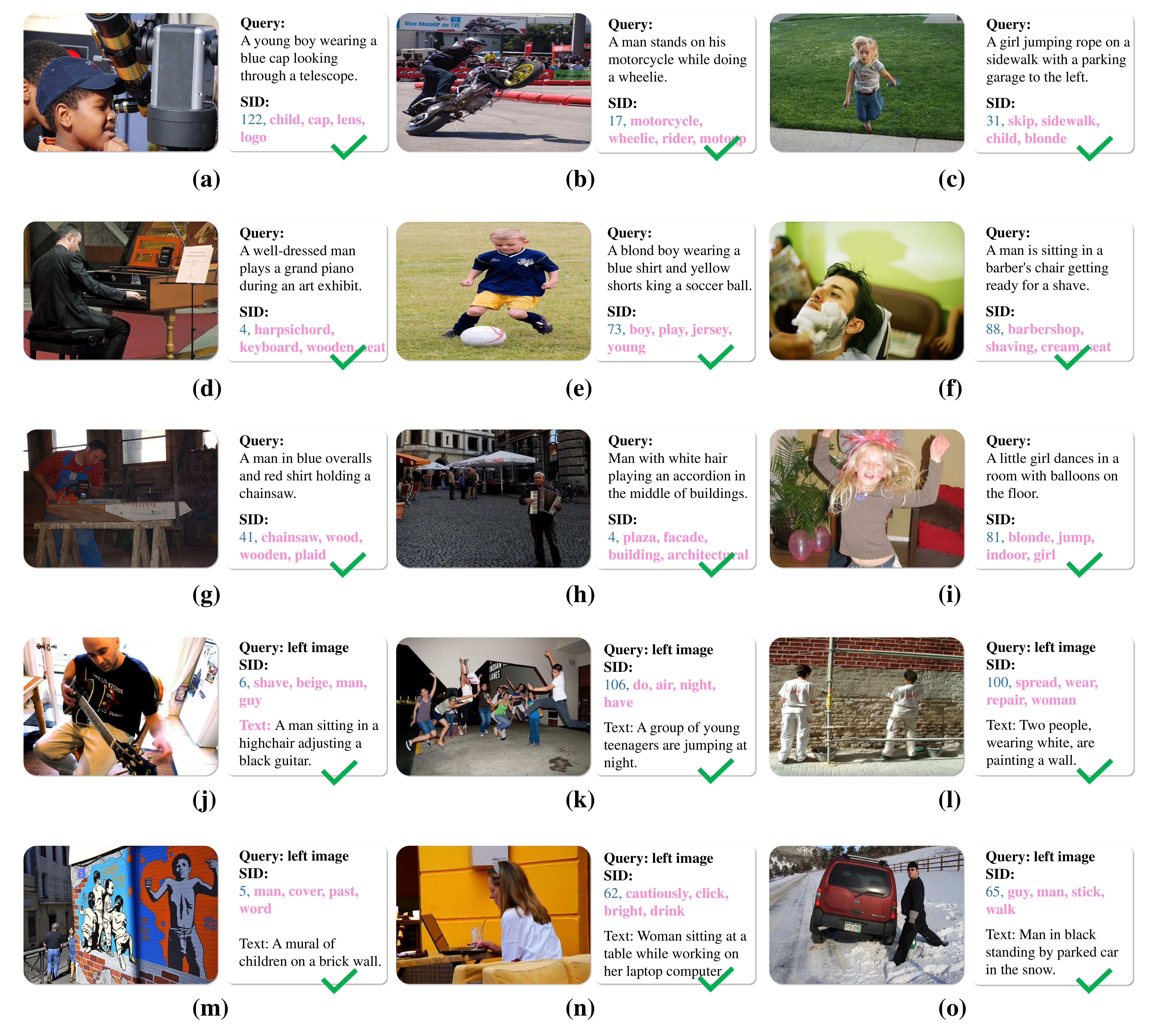}
\caption{Illustration of additional generative retrieval examples. Items from (j) to (o) correspond to the image-to-text retrieval task. The generated SIDs are closely aligned with the corresponding target content. Global IDs within each SID are highlighted in blue, while lexical IDs are marked in pink. Global ID ensures informative and lexical ID ensures discriminative.}
\label{fig:fig_6_sup}
\end{figure*}

\begin{figure*}[ht]
\centering
\includegraphics[width=1.0\textwidth]{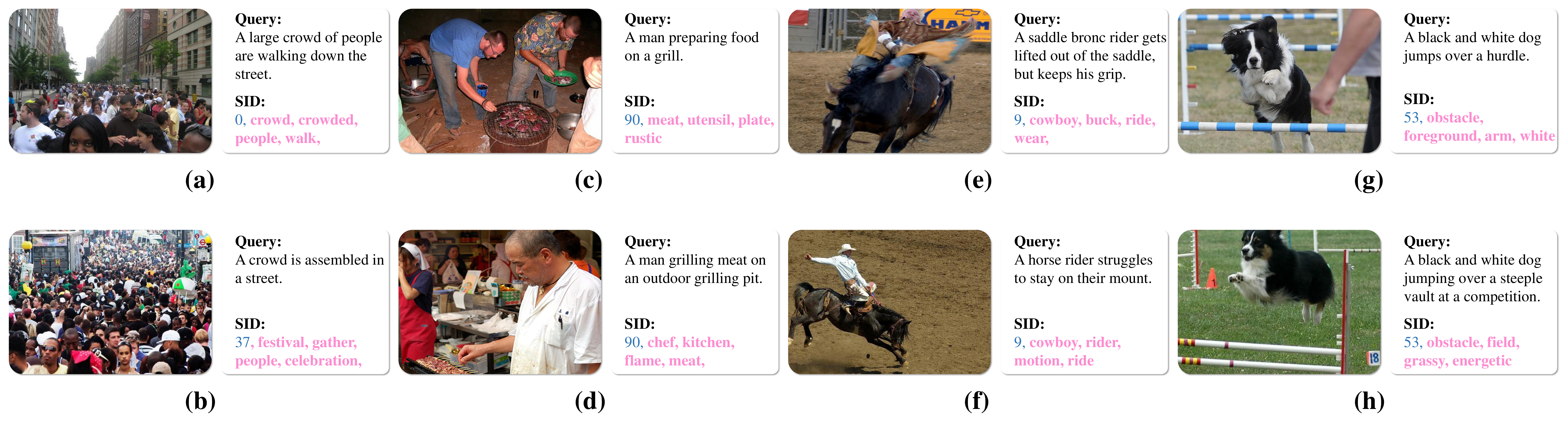}
\caption{Illustration of the false-negative phenomenon. Each column demonstrates two text queries, where each query erroneously generates the SID of the other. Despite this incorrect assignment, the text queries remain semantically relevant to the images associated with the wrong SIDs.}
\label{fig:fig_5_sup}
\end{figure*}

More examples of semantic-enhanced generative cross-modal retrieval are presented in Figure~\ref{fig:fig_6_sup}. These examples demonstrate the effectiveness of our proposed SID in capturing target content. Specifically, the global ID encapsulates the macroscopic information, while the lexical ID focuses on finer details. 
As elaborated in the main paper, global IDs provide informativeness, narrowing the scope for generating subsequent identifier tokens and enabling accurate SID generation. Complementarily, lexical IDs ensure that the SIDs remain discriminative. These examples underscore that the interplay between global and lexical ID ensures SIDs are both informative and discriminative, contributing to the superior performance of the proposed framework as evidenced in Figure~\ref{fig:fig_6_sup}.
Additionally, Figure~\ref{fig:fig_5_sup} provides further examples illustrating the false-negative phenomenon discussed in the main paper. It can be observed that this phenomenon is pervasive across the dataset and can negatively impact the final generative retrieval performance. From a semantic perspective, the generative retrieval results shown in Figure~\ref{fig:fig_5_sup} are successful, as the retrieved images align well with the textual queries in terms of semantic content. However, from the evaluation metric perspective, these results are incorrectly labeled, reflecting a misalignment between semantic relevance and dataset annotations. This phenomenon demonstrates the inadequacy of existing metrics and indirectly illustrates the effectiveness of our framework.

\end{document}